# Ammonium Fluoride as a Hydrogen-disordering Agent for Ice

*Christoph G. Salzmann,*[a] Zainab Sharif,[a] Craig L. Bull,[b] Steven T. Bramwell,[c] Alexander Rosu-Finsen[a] and Nicholas P. Funnell[b]*

[a] Department of Chemistry, University College London, 20 Gordon Street, London WC1H 0AJ, United Kingdom.

[b] ISIS Pulsed Neutron and Muon Source, Rutherford Appleton Laboratory, Harwell Oxford, Didcot OX11 0QX, United Kingdom.

[c] London Centre for Nanotechnology and Department of Physics & Astronomy, University College London, 17-19 Gordon Street, London WC1H 0AJ, United Kingdom.

**Corresponding Author**

* c.salzmann@ucl.ac.uk

**Abstract**

The removal of residual hydrogen disorder from various phases of ice with acid or base dopants at low temperatures has been a focus of intense research for many decades. As an antipode to these efforts, we now show using neutron diffraction that ammonium fluoride ($NH_4F$) is a hydrogen-disordering agent for the hydrogen-ordered ice VIII. Cooling its hydrogen-disordered counterpart ice VII doped with 2.5 mol% $ND_4F$ under pressure leads to a hydrogen-disordered ice VIII with ~31% residual hydrogen disorder illustrating the long-range hydrogen-disordering effect of $ND_4F$. The doped ice VII could be supercooled by ~20 K with respect to the hydrogen-ordering temperature of pure ice VII after which the hydrogen-ordering took place slowly over a ~60 K temperature window. These findings demonstrate that $ND_4F$-doping slows down the hydrogen-ordering kinetics quite substantially. The partial hydrogen order of the doped sample is consistent with the antiferroelectric ordering of pure ice VIII. Yet, we argue that local ferroelectric domains must exist between ionic point defects of opposite charge. In addition to the long-range effect of $NH_4F$-doping on hydrogen-ordered water structures, the design principle of using topological charges should be applicable to a wide range of other 'ice-rule' systems including spin ices and related polar materials.

## Introduction

Water's phase diagram has been explored for more than a century leading to the discovery of 17 crystallographically distinct phases of ice so far.[1-5] The water molecules in the various phases of ice are fully hydrogen-bonded which gives rise to 4-fold connected networks following the 'two-in-two-out' ice rules with respect to the donation and acceptance of hydrogen bonds.[6-8] Based on this building principle, a wide range of network topologies can be observed ranging from the high-symmetry ice I*h* network with only six-membered rings and the highly complex ice V/XIII network to the two interpenetrating individual networks of ice VII/VIII.[6-8]

Within the constraints of the ice rules,[9] the water molecules can display orientational disorder and such phases are categorized as hydrogen-disordered. In fact, all phases of ice that can be crystallized from liquid water display hydrogen disorder as can be seen in Figure 1(a). In the case of ices I*h*, I*sd*, IV, VI, VII and XII, full hydrogen disorder is observed in neutron diffraction in the form of half-occupied deuterium sites.[10-13] Ices III and V on the other hand display partial hydrogen order and some of the fractional occupancies deviate significantly from ½.[13] As required by the third law of thermodynamics, the various hydrogen-disordered phases are expected to transform to their corresponding hydrogen-ordered counterparts upon cooling as long-range orientational order of the water molecules is established. Figure 1(b) shows the crystal structures of the hydrogen-disordered ice VII and its hydrogen-ordered counterpart ice VIII.

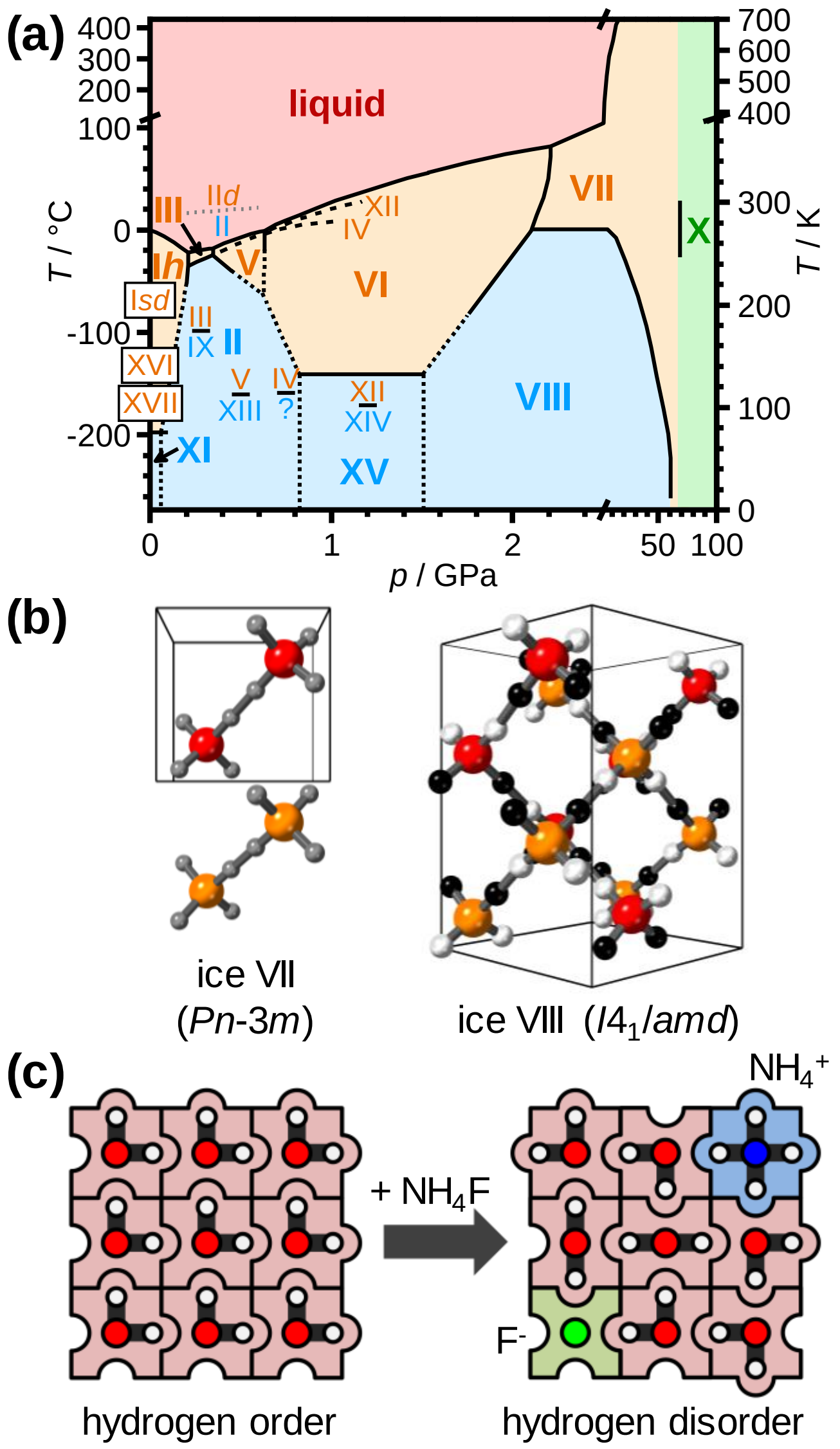

***Figure 1*** *(a) Phase diagram of $H_2O$ including the liquid phase (red) as well as hydrogen-disordered (orange), hydrogen-ordered (blue) and 'polymeric' (green) phases of ice. Metastable melting lines are shown as dashed lines and extrapolated phase boundaries as dotted lines. Stable phases are indicated in bold and metastable phases with a smaller font size. The ice II / ice IId phase boundary was predicted computationally.*[14] *Adapted from ref. 8. (b) Unit cells of ices VII and VIII.*[12, 15-16] *Fully occupied, half occupied and empty hydrogen positions are shown as white, gray and black spheres, respectively. The oxygen atoms of the two individual hydrogen-*

*bonded networks are shown as red and orange spheres. The orientation of the unit cell of ice VIII is such so that the relationship with the unit cell of ice VII can be seen ($a_{VIII} = \sqrt{2}a_{VII}$ and $c_{VIII} = 2a_{VII}$).[15] (c) Schematic illustration of the hydrogen disordering effect of $NH_4F$ doping on the hydrogen-ordered square ice.[8, 17]*

Due to the constraints of the ice rules, the molecular reorientation of water molecules in ice is a difficult process and only possible in the presence of mobile point defects that locally break the ice rules. In the case of ices III and VII, the concentration of intrinsic point defects is sufficient so that the hydrogen-ordering phase transitions to ices IX and VIII can be observed upon cooling.[12, 18] All other hydrogen-disordered phases, including ices I*h*, IV, V, VI and XII, form orientational glasses upon cooling which can display complicated kinetic phenomena around their glass transition temperatures.[5, 19-21] To overcome the problem of glass formation, it has been shown that doping with alkali hydroxides and hydrohalic acids can accelerate molecular reorientations in the hydrogen-disordered phases which then facilitates the hydrogen-ordering processes by introducing mobile defects in the ice-rules network.[1-2, 22-23]

The question of achieving hydrogen order with the help of dopants has been a long-standing and active area of research over many decades.[7-8, 24] To complement these efforts, we have recently begun to investigate the effect of ammonium fluoride ($NH_4F$) doping with respect to inducing hydrogen disorder in hydrogen-ordered phases of ice.[17] The expected hydrogen disordering effect of $NH_4F$ doping is shown schematically in Figure 1(c) for the case of the hydrogen-ordered planar square ice. In ref. 17 we have shown that $NH_4F$-doping leads to the complete disappearance of the hydrogen-ordered ice II from the phase diagram whereas the competing hydrogen-disordered phases, ices I*h*, III, V and VI, are capable of forming solid solutions with $NH_4F$. The disappearance of ice II was explained in terms of an increase in free energy due to the hydrogen disordering which then leads to the formation of the more stable hydrogen-disordered phases. Hence, due to the availability of competing hydrogen-disordered phases, the hydrogen-disordered counterpart of ice II could not be observed in our previous experiments. In this context, it is interesting to note that the hydrogen-disordering phase transition of ice II has been predicted to take place at temperatures where the liquid phase is thermodynamically stable (*cf*. Figure 1(a)) which could explain the experimental difficulties in preparing the hydrogen-disordered counterpart of ice II.[14] As argued in ref. 17, the existence and unusual nature of ice II could explain the host of anomalies observed in water's phase diagram around the pressure range where ice II is the stable phase at low temperatures.

Here we investigate the effect of $NH_4F$-doping on the ice VII $\rightarrow$ ice VIII phase transition using *in-situ* neutron diffraction. Since the hydrogen-disordered ice VII is easily accessible, it should be possible to monitor the effect of $NH_4F$ on the formation of ice VIII directly without the danger of transformations to competing phases as previously observed. The high-symmetry structures of ices VII and VIII, which facilitates the analyses of the diffraction data, make them perfectly suited for this study.

**Experimental Methods**

Deuterated ammonium fluoride ($ND_4F$) was obtained by dissolving $NH_4F$ (99.99% trace metal basis) in an excess of deuterated water (99.9% D) followed by complete evaporation of the liquid under dinitrogen gas. To ensure complete deuteration, these steps were repeated twice. The resulting dry $ND_4F$ powder was stored in a desiccator and used to prepare a 2.5 mol% solution in $D_2O$. All $ND_4F$ solutions were prepared and handled in polyethylene containers.

For the first neutron diffraction experiment, some of the $ND_4F$ solution was frozen by dropping onto a cold surface at liquid nitrogen temperature and ground to a fine powder using a porcelain pestle and mortar under liquid nitrogen. The ice powder was then transferred into a precooled null-scattering encapsulated TiZr gasket mounted on zirconia-toughened alumina anvils of a Paris-Edinburgh press and compacted as firmly as possible. Before closing the anvil cell, a small lead bead was placed in the center of the ice sample to be used as a pressure calibrant. A load of 8 tonnes was then applied to seal the cell within a precooled Paris-Edinburgh setup, which was mounted on the PEARL beamline[25] at the ISIS Neutron and Muon Source, Didcot, UK. The applied load was then increased in steps of 5 tonnes up to 58 tonnes. The temperature of the press was controlled with a heating element and a reservoir of liquid nitrogen at the bottom of the containment vessel.

In a separate experiment, a pure $D_2O$ ice VIII sample was prepared by freezing ice VII under pressure from the liquid and cooling to 250 K. All neutron diffraction patterns were collected at 90° scattering angle and analyzed using the GSAS software.[26] The pressure of the sample was determined by refining the lattice constant of lead and then using its previously reported equation of state.[27] The estimated errors in the obtained pressures are ±0.19 GPa.

## Results and Discussion

For the preparation of $ND_4F$-doped ice VII, doped ice I*h* was compressed to 4.61 GPa while simultaneously increasing the temperature from 190 to 290 K as shown in Figure 2(a). The neutron diffraction patterns collected during the compression are shown in Figure 2(b) and displayed a cascade of phase transitions from ice I*h* → ice III → ice V → ice XII → ice VI → ice VIII / VII. Remarkably, and in line with our previous study,[17] the formation of ice II was fully suppressed due to the presence of the $ND_4F$.

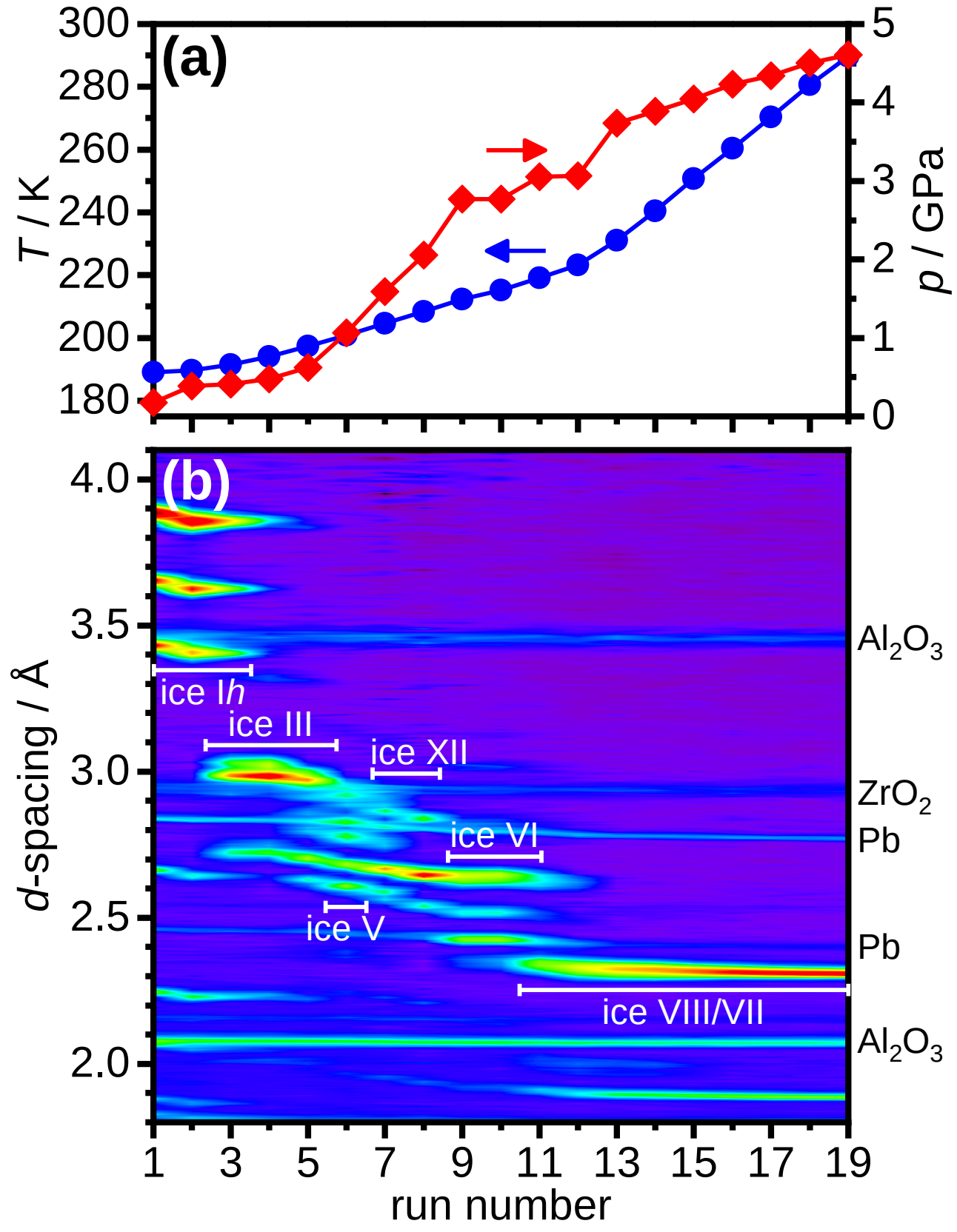

***Figure 2*** *Preparation of 2.5 mol% $ND_4F$ doped ice VII. (a) Increases in temperature and pressure during the compression experiment. (b) Corresponding neutron diffraction patterns displayed as a contour plot using the square roots of the intensities to emphasize weaker features. The ranges of existence of the various ice polymorphs are highlighted as well as the Bragg peaks due to the anvil and the Pb pressure calibrant.*

At 290 K and 4.61 GPa, a high-quality neutron diffraction pattern was collected as shown in Figure 3(a). The pattern could be fitted well with the hydrogen-disordered *Pn-3m* ice VII structural model with the nitrogen atoms of $ND_4^+$ and the $F^-$ anions positioned on the oxygen site, and the deuterium atoms of $ND_4^+$ located at the deuterium sites of $D_2O$ (*cf*. Table 1). This suggests that the $ND_4^+$ and $F^-$ ions are randomly substituted into the hydrogen-bonded networks of ice VII, which results in a half-occupied deuterium site on average as previously observed for $ND_4F$-doped ice I*h*.[17] The presence of an $ND_4^+$ cation on a given oxygen site increases the occupancies of all four nearby deuterium sites. However, this increase is cancelled by the equally probable presence of a fluoride anion.

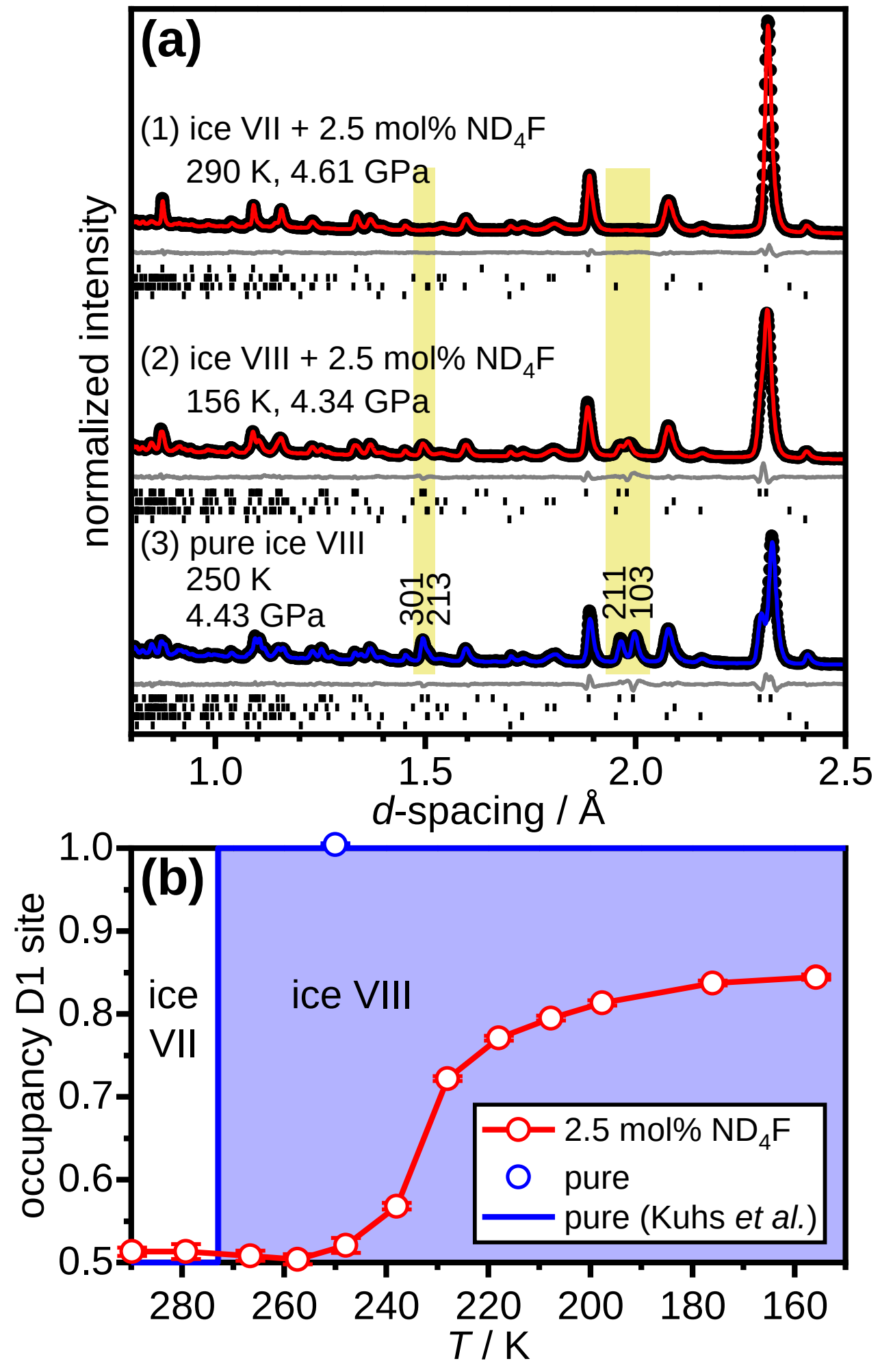

***Figure 3*** *(a) Rietveld analysis of (1) $ND_4F$-doped ice VII at 290 K and 4.61 GPa, (2) $ND_4F$-doped ice VIII at 156 K and 4.34 GPa, and (3) pure ice VIII at 250 K and 4.43 GPa. The experimental data are shown as black circles, red or blue lines are the Rietveld fits, and gray lines are the differences between the experimental data and the Rietveld fits downshifted for*

*clarity. The tick marks under each of the diffraction patterns are from top to bottom: ice VII/VIII, $ZrO_2$, $Al_2O_3$ and Pb. The yellow-shaded areas highlight the d-spacing ranges where Bragg peaks characteristic for ice VIII are observed. (b) Refined occupancies of the D1 deuterium site of the $ND_4F$-doped (red) and pure ice VII/VIII (blue) as a function of temperature.*

In the next step, the $ND_4F$-doped ice VII sample was cooled at 0.3 K $min^{-1}$ from 290 to 156 K while keeping the load on the pressure cell constant. As shown in Figure 4(a), the pressure decreased slightly to 4.34 GPa during the essentially isochoric cooling process. The corresponding neutron diffraction patterns in Figure 4(b) show that the ice VII persisted down to about 240 K where Bragg peaks characteristic for ice VIII started to appear at around 1.49 and 1.97 Å. A high-quality diffraction pattern of the $ND_4F$-doped sample at 156 K is shown in Figure 3(a) and compared with a diffraction pattern of pure ice VIII at 250 K. It can be seen that the intensities of the Bragg peaks characteristic for ice VIII display weaker intensities for the $ND_4F$-doped sample and the splitting of most intense feature at about 2.3 Å, which is one of the hallmark features of the ice VII $\rightarrow$ VIII phase transition, is less pronounced. These observations already suggest that the $ND_4F$-doped ice VIII is less hydrogen-ordered than the fully hydrogen-ordered pure ice VIII.

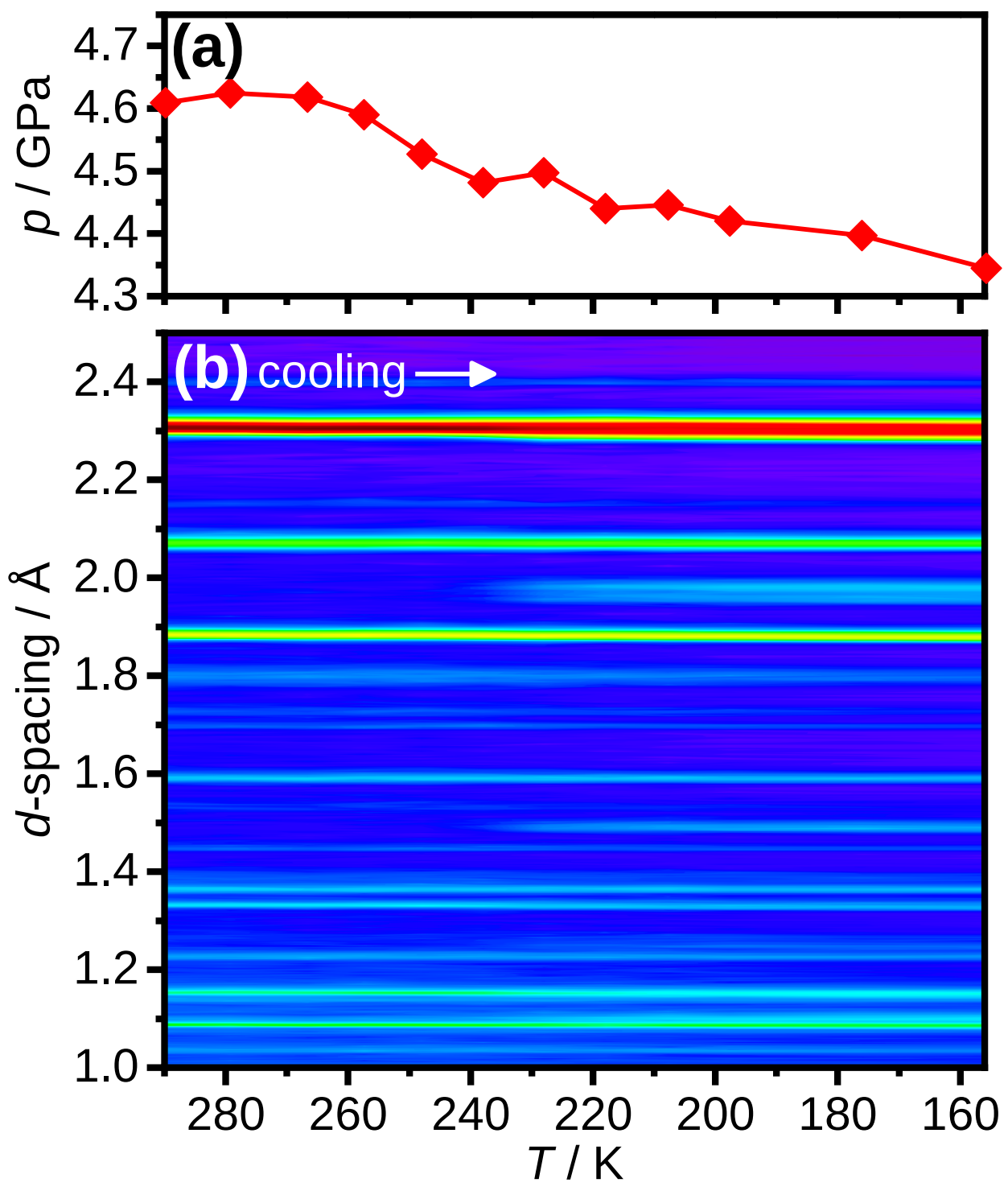

***Figure 4*** *Cooling 2.5 mol% $ND_4F$-doped ice VII under pressure at 0.3 K $min^{-1}$. (a) Changes in pressure upon cooling. (b) Corresponding neutron diffraction patterns displayed as a contour plot using the square roots of the intensities to emphasize weaker features.*

The Rietveld refinement of the diffraction pattern of $ND_4F$-doped ice VIII at 156 K revealed that its two deuterium sites indeed display residual hydrogen disorder (*cf.* Table 2). The fractional occupancies of the two sites were determined as 0.844 ± 0.002 and 0.156 ± 0.002, respectively. This means that the doping with 2.5 mol% $ND_4F$ leads to an average hydrogen-disordering effect, *i.e.* deviation of the occupancies from either one or zero, of 0.156. Considering that an occupancy of ½ corresponds to full hydrogen disorder, the residual hydrogen disorder can be calculated as 0.156 / 0.5 which equals 31.2%.

Due to the substitution of $ND_4^+$ onto an oxygen site, the occupancies of the neighboring deuterium sites within hydrogen-ordered ice VIII must be greater than 0.024 at a $ND_4F$ concentration of 2.5 mol% since the four deuterium atoms of $ND_4^+$ inevitably make the neighboring deuterium sites occupied. Equally, the substitution with $F^-$ anions means that the maximal value of the occupancies of the neighboring deuterium sites cannot be above 0.977. The fact that the observed average occupancies of 0.844 and 0.156 deviate much more considerably from full hydrogen order illustrates that the $ND_4F$ doping of ice VIII has a long-range hydrogen disordering effect that goes far beyond the first-neighbor effects of the substitution.

In ref. 17, we described the effect of $NH_4F$ doping in terms of the ions acting as 'topological charges' that disrupt the orientational order of the water molecules over long distances along a system of 'flux tubes' in the hydrogen atoms' displacements between the $NH_4^+$ and $F^-$ defects. In the case of $NH_4F$-doped ice II, the increase in free energy was sufficient so that ice II disappeared altogether from the phase diagram, and ices III and V formed instead.[17] For this reason, the hydrogen-disordering effect of $NH_4F$ could not be directly proven in our previous work. However, since the hydrogen-disordered ice VII is easily accessible, the hydrogen-disordering effect of $NH_4F$ could now be demonstrated for ice VIII. The hydrogen-disordering effect of $NH_4F$ doping is of course what chemical intuition would suggest (*cf*. Figure 1(c)). Yet, it also needs to be emphasized that the flux tubes between the ions can only develop if a hydrogen-ordered phase of ice is free of other point defects such as ionic ($H_3O^+$ and $OH^-$) or Bjerrum (O-H H-O and O…O) defects. The presence of such defects would allow screening of the topological charges and so inevitably disrupt the flux tubes. In the case of ice VIII, the absence of dielectric relaxation suggests a very low level or even absence of such defects,[28] which are certainly far below the detection limit of diffraction.[12, 15] But also, the hydrogen-disordering effect of $NH_4F$ doping presented here should also be seen as evidence for low levels of such defects.

For pure ice VII, it is well-known that the hydrogen ordering to the antiferroelectric ice VIII takes place at around 0°C up to ~10 GPa (*cf*. Figure 1(a)) and with a hysteresis of only about 5.5 K.[28] So, compared to other hydrogen-ordering phase transitions,[1, 5, 20, 22-23, 29-31] it is quite sharp and in fact the only one that proceeds from full hydrogen-disorder to full hydrogen-order according to diffraction.[12, 15-16] In line with previous studies, our pure $D_2O$ ice VIII displayed complete hydrogen order at 250 K and 4.43 GPa as indicated by the blue data point in Figure 3(b).

To investigate the only partial hydrogen ordering upon cooling the $ND_4F$-doped sample in more detail, all diffraction patterns starting from ice VII at 290 K down to 156 K were analyzed on the basis of the $I4_1/amd$ structural model of ice VIII, which can be used to describe the higher symmetry ice VII but not the other way around. The refined fractional occupancies of the D1 deuterium site are shown as a function of temperature in Figure 3(b). Since the D1 and D2 deuterium positions are located along the same type of hydrogen bond, the fractional occupancies of D2 would be simply 1 minus the occupancy of the D1 site. Remarkably and in contrast to pure ice VII,[12] $ND_4F$-doped ice VII can be supercooled down to 250 K where slow hydrogen ordering sets in. At this point, it is important to recall that a slow cooling rate of 0.3 K $min^{-1}$ was used in our experiment. For pure $D_2O$ ice VII, it has been noted that the sharp phase transition to ice VIII cannot be suppressed upon cooling even if the sample is cooled quite rapidly at ~20 K $min^{-1}$.[32] For the $ND_4F$-doped sample, the hydrogen ordering takes place over an about 60 K range and seems to level off below 180 K. These findings illustrate that in addition to inducing hydrogen disorder in the final state at low temperatures, $NH_4F$-doping influences the kinetics of the hydrogen ordering as well. The hydrogen ordering of ice VII, which takes place within a very narrow temperature window for the pure material, is significantly slowed-down by the presence of the dopant. It is generally acknowledged that the mechanism of hydrogen ordering relies on mobile point defects within the hydrogen-disordered phases which leave behind a trail of water molecules with changed orientations.[8] For such migrating defects, the $NH_4^+$ or $F^-$ ions represent 'dead ends' beyond which they cannot travel. So, instead of moving freely like in the pure ice phases, the defect trails in $NH_4F$-doped ice can be envisaged as being confined between the extrinsic ionic defects. Effectively, the $NH_4F$-doping leads to a reduction of the fractal dimension of the defect trails and this effect is expected to scale with the concentration of the $NH_4F$ dopant.

At ambient pressure, $NH_4F$ is soluble in ice I*h* across a wide composition range[33-34] and it has been shown to be soluble in clathrate hydrates[35-36] as well as in ice III.[17] This is remarkable since most other ionic species display very low solubilities in ice.[37] The hydrogen-disordering effect of $NH_4F$ shows that it soluble in ices VII and VIII as well. In fact, $NH_4F$-III, which is the stable phase of $NH_4F$ above ~1 GPa,[38-39] is thought to be isostructural with ice VII/VIII.[40] The formation of solid solutions with ice VII/VIII is therefore probably to be expected.

Furthermore, the solubility of $NH_4F$ in ice VII/VIII is in line with recent studies showing that lithium halides are soluble in ice VII at ~1:6 LiX:$H_2O$ molar ratios.[41-42] This 'salty' ice VII has been shown to be highly disordered and a hydrogen-ordering phase transition to ice VIII could not be observed upon cooling. However, this is perhaps not surprising considering that the ratio of ionic species to water was 1:3. In our study, a much lower concentration of 2.5 mol% $ND_4F$ was used corresponding to a 1:21 ratio of ionic species to water which highlights the long-range hydrogen-disordering effect of the $NH_4F$ doping. In addition to the lithium halides, sodium and magnesium halides also seem to be soluble in ice VII although at much lower concentrations.[43-46]

In addition to the 'chemical' strategy for hydrogen-disordering ice VIII presented here, it is interesting to note that hydrogen-disordered ice VII, or at least a material closely related to ice VII called ice VII', can be obtained by compression of ice VI or high-density amorphous ice below 95 K.[32, 47] Unlike our doped hydrogen-disordered ice VIII, ice VII' is metastable and transforms to the stable ice VIII upon heating under pressure.[32, 47]

Most recently, it was attempted to induce ferroelectric ordering in ice VII using an external electric field[48] and it was argued that the ferroelectric domains could have $P4_2nm$ space group symmetry.[49] In this context, it is interesting to note that the partial hydrogen order in the $ND_4F$-doped ice VIII sample could be described with the antiferroelectric $I4_1/amd$ structural model as shown in Figure 3(a). Nevertheless, the substitution of $NH_4^+$ and $F^-$ ions into an antiferroelectric hydrogen-ordered structure is expected to induce at least some degree of local ferroelectric ordering between ions of opposite charge. This effect arises from the orientation flipping of the water molecules along the two 'flux tubes' between the ions discussed earlier. The Coulomb field associated with the ions is expected to play a very minor part in this process apart from in the sense that it affects correlation between ions through screening. Since the ions are randomly distributed within ice VII/VIII, the net effect of the local ferroelectric ordering is expected to cancel and give an overall average structure that is antiferroelectric.

**Conclusions**

We have shown that $NH_4F$ is a hydrogen-disordering agent for hydrogen-ordered ice. The molecular mechanism relies on the disruption of the 'two-in-two-out' rule at the sites of the ionic point defects which leads to two long-range hydrogen-disordering trails between ions of opposite charge. In addition to providing a nice example of crystal engineering for the hydrogen-ordered bulk phases of ice, it seems likely that $NH_4F$ doping could also be used to influence the hydrogen-ordering phase transitions in organic and inorganic hydrates.[50-51] But also, the concept of introducing 'four-in-zero-out' and 'zero-in-four-out' point defects may be applicable to other systems following the ice rules such as spin ices,[52-53] spin-ice related cyanide materials[54] and certain types of polymerization reactions.[55] Finally, it can be speculated that local effects of $NH_4F$ doping may be connected with its inhibiting effect on methane clathrate hydrate formation[36] and it is worth mentioning that ammonium salts have recently been shown, in stark contrast to the colligative effects of other salts, to raise the heterogeneous ice nucleation temperatures.[56]

**Acknowledgment**

Funding is acknowledged from the Royal Society for a University Research Fellowship (UF100144), the UCL Chemistry Department for a DTA studentship, and the European Research Council under the European Union's Horizon 2020 research and innovation programme (grant agreement No 725271). We also thank ISIS for granting access to the PEARL beamline and C. Ridley for help with the neutron diffraction experiment. The raw data of this experiment is available at https://data.isis.stfc.ac.uk/doi/investigation/86391366.

***Table 1.*** *Fractional atomic coordinates, fractional occupancies and isotropic atomic-displacement parameters ($U_{iso}$) of $D_2O$ Pn-3m ice VII with 2.5 mol% $ND_4F$ at 290 K and 4.61 GPa. The lattice constants are: a = b = c = 3.26914(7) Å. Numbers in parentheses are statistical errors of the last significant digit of refined quantities. The occupancies related to the oxygen, nitrogen and fluorine atoms were calculated from the $ND_4F$ concentration of the initial solution.*

| atom label | atom type (Wyckoff) | *x* | *y* | *z* | occupancy | $U_{iso}$*100 |
|---|---|---|---|---|---|---|
| O1 | O (2a) | ¼ | ¼ | ¼ | 0.9512 | 1.640(29) |
| D1 | D (8e) | 0.5857(1) | 0.5857(1) | 0.5857(1) | ½ | 2.273(31) |
| N1 | N (2a) | ¼ | ¼ | ¼ | 0.0244 | 1.640(29) |
| F1 | F (2a) | ¼ | ¼ | ¼ | 0.0244 | 1.640(29) |

***Table 2.*** *Fractional atomic coordinates, fractional occupancies and isotropic atomic-displacement parameters ($U_{iso}$) of $D_2O$ $I4_1$/amd ice VIII with 2.5 mol% $ND_4F$ at 156 K and 4.34 GPa. The lattice constants are: a = b = 4.59104(17) Å, c = 6.5809(4) Å. Numbers in parentheses are statistical errors of the last significant digit of refined quantities. The occupancies related to the oxygen, nitrogen and fluorine atoms were calculated from the $ND_4F$ concentration of the initial solution.*

| atom label | atom type (Wyckoff) | *x* | *y* | *z* | occupancy | $U_{iso}$*100 |
|---|---|---|---|---|---|---|
| O1 | O (8e) | 0 | ¼ | 0.1124(4) | 0.9512 | 0.76(4) |
| D1 | D (16h) | 0 | 0.4214(5) | 0.2005(3) | 0.844(2) | 1.52(4) |
| D2 | D (16h) | 0 | 0.091(2) | 0.709(2) | 0.156(2) | 1.52(4) |
| N1 | N (8e) | 0 | ¼ | 0.1124(4) | 0.0244 | 0.76(4) |
| F1 | F (8e) | 0 | ¼ | 0.1124(4) | 0.0244 | 0.76(4) |